\documentclass[sigconf,natbib=true]{acmart}

\usepackage{amsmath}

\usepackage{amssymb}
\usepackage{algorithm}
\usepackage{algorithmic}
\usepackage{graphicx}
\usepackage{booktabs}
\usepackage{xcolor}
\usepackage{hyperref}
\usepackage{microtype}
\usepackage{tikz}
\usetikzlibrary{shapes.geometric, arrows.meta, positioning}

\setcopyright{none}             
\acmConference[]{}{}{}
\acmDOI{}
\acmISBN{}
\copyrightyear{}

\title{Aperon Technical Report: Hierarchical No-Pointer Tangent-Local Search for High-Dimensional Approximate Nearest Neighbors}

\author{Yong Fu}
\affiliation{%
  \institution{Substratum Labs}
  \country{}
}
\email{yong@substratumlabs.ai}

\begin{abstract}
We present HNTL (Hierarchical No-pointer Tangent-Local), the core vector indexing and candidate generation framework of the \emph{Aperon} vector memory system. Proximity graphs (e.g., HNSW) incur a heavy pointer tax in memory overhead and induce irregular memory accesses that stall CPU pipelines. HNTL resolves this by partitioning the high-dimensional space into local, coherent \emph{grains}, representing vectors as low-dimensional coordinates on local tangent spaces, and scanning them sequentially using a pointerless Block-SoA (Structure-of-Arrays) layout. 
On anisotropic manifold data ($d=768$, $N=10{,}000$), local PCA captures $96.3\%$ of the variance, allowing HNTL to achieve a final Rerank Recall@10 of 1.0000 with a candidate pool size of only $C=20$ vectors. Hardware profiling via Apple \texttt{kperf} CPU Performance Monitoring Unit (PMU) counters demonstrates a 3.61x speedup ($4.137$\,ns/vector vs. $14.951$\,ns/vector) for our NEON auto-vectorized Rust Block-SoA scan engine over standard pointer-chasing graph traversals, driven by a $3.59\times$ IPC (Instructions Per Cycle) and near-zero L1/L2 data cache misses.

\end{abstract}


\begin{document}
\pagestyle{plain}
\makeatletter
\let\ps@firstpage\ps@plain
\makeatother
\maketitle

\section{Introduction \& Background}
\label{sec:intro}

High-dimensional approximate nearest neighbor (ANN) search is a core primitive in modern machine learning systems, including retrieval-augmented generation (RAG) and long-running cognitive agents. Traditional approaches range from locality-sensitive hashing (LSH)~\cite{indyk1998approximate} to product quantization (PQ)~\cite{jegou2011product} and its optimized variants (OPQ)~\cite{ge2013optimized}. In production, standard vector databases rely on proximity graphs (e.g., HNSW~\cite{malkov2018efficient} or graph implementations in FAISS~\cite{johnson2019billion,douze2024faiss}), which chase memory pointers during traversal. This design has two critical flaws for agent environments:
\begin{enumerate}
  \item \textbf{The Pointer Tax}: Graph links frequently exceed raw vector data in size, requiring substantial DRAM.
  \item \textbf{Cognitive Impedance}: Long-running agents require transactional properties like instant zero-copy branching (counterfactual reasoning), snapshots, and unified mixed-recall (vector similarity combined with temporal, spatial, and symbolic filters).
\end{enumerate}
To resolve these limitations, the production core of the Aperon database~\cite{aperon2026} is implemented in Rust to guarantee memory safety, concurrency, and high-performance execution. 

Aperon resolves the \emph{pointer tax} by introducing the HNTL (Hierarchical No-pointer Tangent-Local) framework. Mathematically, HNTL abandons graph links completely by partitioning the vector space into localized grains and projecting vectors onto low-dimensional local tangent spaces. Algorithmically, searching for nearest neighbors within a grain is transformed into a sequential local scan rather than a pointer-chasing graph traversal. Physically, this scan is implemented using a pointerless, cache-aligned Block-SoA memory layout, enabling the CPU to run SIMD-accelerated execution with near-zero cache misses.

To resolve \emph{cognitive impedance}, HNTL's localized grain structure is mapped directly to Aperon's immutable Memory SSTable (Log-Structured Memory Segment) architecture. Because grains are self-contained and geographically isolated, updating or adding vectors does not trigger global graph re-wiring. This allows the database to support instant, zero-copy branching (via copy-on-write segment forks for parallel counterfactual simulations) and simplifies mixed-recall queries, as symbolic or temporal filters can be checked in-situ within the sequential scan loop.

Specifically, the system is structured around a hierarchical routing plane and a dual-mode query planner:
\begin{itemize}
  \item \textbf{Mode A (Self-Contained)}: Performs online vector reconstruction directly in DRAM from local projections, bypassing disk reads completely.
  \item \textbf{Mode B (Tiered Filter)}: Uses a hot DRAM-resident manifold index ($<8\%$ HNSW footprint) as a candidate generator, loading raw vectors from cold storage (SSD/mmap) for exact re-ranking.
\end{itemize}

\section{System Architecture \& Design}
\label{sec:design}

The Aperon database engine is structured around three primary layers to enable high-efficiency agent memory storage and retrieval (see Figure~\ref{fig:architecture}):
\begin{itemize}
  \item \textbf{Log-Structured SSTable Layer}: Consists of immutable memory segments on disk or memory-mapped files. Each segment groups raw records and their index access paths in contiguous, pointerless \textbf{Block-SoA} structures to enable hardware-friendly linear scanning.
  \item \textbf{Hierarchical Routing Plane}: A coarse centroid index (routing table) that maps queries to matching candidate grains, pruning grains that fail a query-specific envelope filter.
  \item \textbf{Dual-Mode Query Planner}: Dynamically schedules query tasks. Under \emph{Mode A}, it reconstructs approximate vectors directly from tangent space coordinates. Under \emph{Mode B}, it routes queries to candidates before performing exact L2 re-ranking.
\end{itemize}

\subsection{Query Execution Lifecycle}
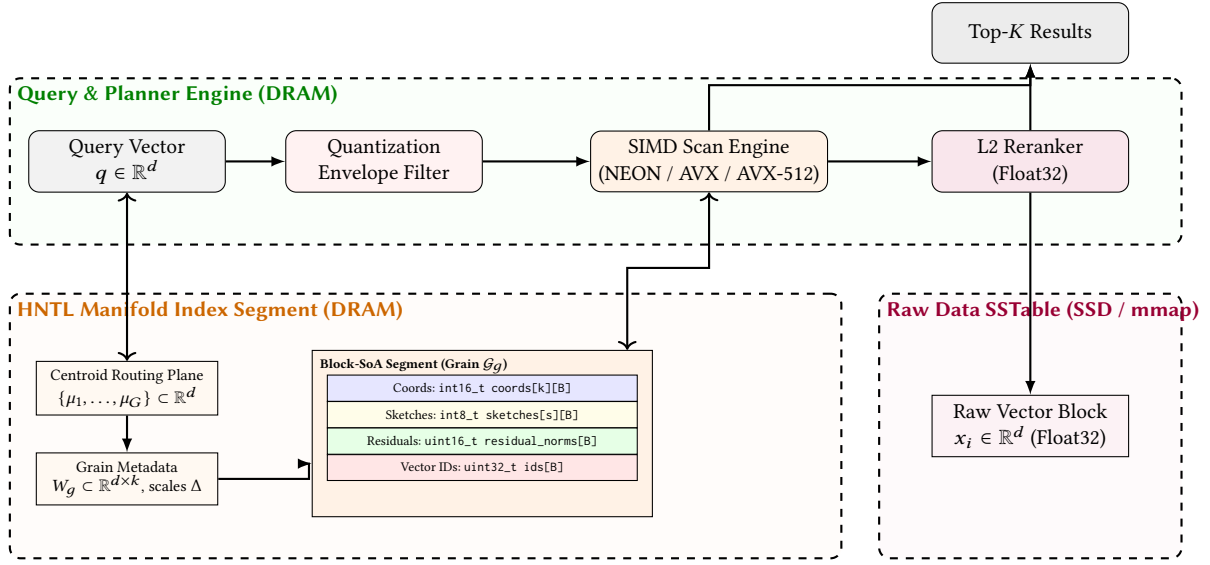
\begin{figure*}[t]
\centering
\begin{tikzpicture}[
    layer/.style={draw, dashed, thick, rounded corners, fill=gray!2, inner sep=0.4cm},
    box/.style={draw, rectangle, rounded corners, minimum width=2.6cm, minimum height=0.8cm, align=center, fill=blue!5, font=\small},
    membox/.style={draw, rectangle, minimum width=2.4cm, minimum height=0.6cm, align=center, fill=orange!5, font=\scriptsize},
    dbbox/.style={draw, rectangle, minimum width=2.6cm, minimum height=0.7cm, align=center, fill=purple!5, font=\small},
    arrow/.style={-latex, thick}
]
    
    \node[layer, minimum width=15.5cm, minimum height=2.2cm, fill=green!2] (engine_layer) at (0, 3.5) {};
    \node[anchor=north west, font=\sffamily\bfseries\small, text=green!50!black] at (engine_layer.north west) {Query \& Planner Engine (DRAM)};
    
    \node[layer, minimum width=11.0cm, minimum height=3.5cm, fill=orange!2] (dram_layer) at (-2.25, 0) {};
    \node[anchor=north west, font=\sffamily\bfseries\small, text=orange!80!black] at (dram_layer.north west) {HNTL Manifold Index Segment (DRAM)};
    
    \node[layer, minimum width=4.0cm, minimum height=3.5cm, fill=purple!2] (ssd_layer) at (5.75, 0) {};
    \node[anchor=north west, font=\sffamily\bfseries\small, text=purple!80!black] at (ssd_layer.north west) {Raw Data SSTable (SSD / mmap)};
    
    \node (query) [box, fill=gray!10] at (-6.2, 3.5) {Query Vector\\$q \in \mathbb{R}^d$};
    \node (filter) [box, fill=red!5] at (-2.8, 3.5) {Quantization\\Envelope Filter};
    \node (scan_kernel) [box, fill=orange!10] at (1.5, 3.5) {SIMD Scan Engine\\(NEON / AVX / AVX-512)};
    \node (reranker) [box, fill=purple!10] at (5.75, 3.5) {L2 Reranker\\(Float32)};
    
    \node (centroids) [membox] at (-6.2, 0.5) {Centroid Routing Plane\\$\{\mu_1, \dots, \mu_G\} \subset \mathbb{R}^d$};
    \node (metadata) [membox] at (-6.2, -0.7) {Grain Metadata\\$W_g \subset \mathbb{R}^{d \times k}$, scales $\Delta$};
    
    \node[membox, minimum width=4.5cm, minimum height=2.2cm, fill=orange!10] (blocksoa) at (-1.5, -0.1) {};
    \node[anchor=north west, font=\tiny\bfseries] at (blocksoa.north west) {Block-SoA Segment (Grain $\mathcal{G}_g$)};
    
    \node[draw, rectangle, minimum width=4.1cm, minimum height=0.3cm, fill=blue!10, font=\tiny] (coords) at (-1.5, 0.5) {Coords: {\tt int16\_t coords[k][B]}};
    \node[draw, rectangle, minimum width=4.1cm, minimum height=0.3cm, fill=yellow!10, font=\tiny] (sketches) at (-1.5, 0.15) {Sketches: {\tt int8\_t sketches[s][B]}};
    \node[draw, rectangle, minimum width=4.1cm, minimum height=0.3cm, fill=green!10, font=\tiny] (residuals) at (-1.5, -0.2) {Residuals: {\tt uint16\_t residual\_norms[B]}};
    \node[draw, rectangle, minimum width=4.1cm, minimum height=0.3cm, fill=red!10, font=\tiny] (ids) at (-1.5, -0.55) {Vector IDs: {\tt uint32\_t ids[B]}};
    
    \node (raw_vectors) [dbbox] at (5.75, 0) {Raw Vector Block\\$x_i \in \mathbb{R}^d$ (Float32)};
    
    \draw[arrow] (query) -- (filter);
    \draw[arrow] (filter) -- (scan_kernel);
    \draw[arrow] (scan_kernel) -- (reranker);
    
    \draw[arrow, <->] (query.south) -- (centroids.north);
    \draw[arrow] (centroids.south) -- (metadata.north);
    \draw[arrow] (metadata.east) -- ++(1.2,0) |- (blocksoa.190);
    
    \draw[arrow, <->] (scan_kernel.south) -- ++(0, -1.0) -| (blocksoa.30);
    \draw[arrow] (reranker.south) -- (raw_vectors.north);
    
    \node (output) [box, fill=gray!15] at (5.75, 5.2) {Top-$K$ Results};
    \draw[arrow] (reranker.north) -- (output.south);
    \draw[arrow] (scan_kernel.north) |- ++(1.0, 0.6) -| (output.south);
    
\end{tikzpicture}
\caption{Aperon HNTL System Component Architecture (DRAM vs. SSD Memory layout).}
\label{fig:architecture}
\end{figure*}

When a query vector $q \in \mathbb{R}^d$ is submitted to the HNTL engine, it executes through the following stages (as shown in Figure~\ref{fig:architecture}):
\begin{enumerate}
  \item \textbf{Centroid Routing}: The engine calculates the Euclidean distance between $q$ and all grain centroids $\mu_g$ in the global routing plane, selecting the top-$P$ (e.g., $nprobe$) closest candidate grains.
  \item \textbf{Manifold Projection \& Filtering}: For each candidate grain, $q$ is projected onto the local PCA basis to produce subspace coordinate $z_{q, g}$ and residual $r_q$. The Quantization Envelope Filter checks for coordinate saturation; if coordinates clip across more than $25\%$ of dimensions, the grain is pruned.
  \item \textbf{Block-SoA Scanning}: The SIMD scan engine sequentially scans the remaining candidate grains in DRAM using the Block-SoA layout, computing distance approximations using integer register math.
  \item \textbf{Query Planner Resolution}:
  \begin{itemize}
    \item \textbf{Mode A (Self-Contained)}: The approximate nearest neighbors are returned directly from the DRAM-based scan results.
    \item \textbf{Mode B (Tiered Filter)}: The query planner fetches the original high-dimensional vectors corresponding to the DRAM candidate set from cold storage (SSD or memory-mapped segments) and performs exact float32 L2 re-ranking.
  \end{itemize}
\end{enumerate}

\subsection{HNTL Subspace Projection}
The core idea of HNTL (Hierarchical No-pointer Tangent-Local) is that while high-dimensional vector embeddings generally occupy a huge global space, vectors in a localized neighborhood (a \emph{grain}) tend to lie on or near a low-dimensional flat surface (a tangent space). By exploiting local geometric coherence, HNTL represents vectors using low-dimensional subspace coordinates, eliminating the need to store and scan full-dimensional data or traverse pointer-heavy index graphs.

\begin{figure}[h]
\centering
\begin{tikzpicture}[
    node distance=0.8cm,
    block/.style={draw, rectangle, rounded corners, minimum width=3.5cm, minimum height=0.6cm, align=center, fill=blue!5, font=\small},
    arrow/.style={-latex, thick}
]
    \node (input) [block, fill=gray!10] {Input Vector $v \in \mathbb{R}^d$\\ (Data $x_i$ or Query $q$)};
    \node (center) [block, below=of input] {1. Mean Centering\\ $v' = v - \mu_g$};
    \node (proj) [block, below=of center] {2. Subspace Projection\\ $z = W_g^\top v' \in \mathbb{R}^k$};
    \node (recon) [block, below=of proj] {3. Subspace Reconstruction\\ $\tilde{v} = W_g z \in \mathbb{R}^d$};
    \node (res) [block, below=of recon] {4. Residual Calculation\\ $e = v' - \tilde{v} \in \mathbb{R}^d$};
    \node (norm) [block, below=of res] {5. Residual Norm\\ $r = \|e\|_2^2 \in \mathbb{R}$};
    \node (quant) [block, below=of norm, fill=orange!5] {6. 16-bit Quantization\\ $z \to \hat{z}$ (int16), $r \to \hat{r}$ (uint16)};
    \node (output) [block, below=of quant, fill=green!5] {Output: $\hat{z}$ (int16) \& $\hat{r}$ (uint16)\\ (Stored in Block-SoA or scanned)};
    
    \draw [arrow] (input) -- (center);
    \draw [arrow] (center) -- (proj);
    \draw [arrow] (proj) -- (recon);
    \draw [arrow] (recon) -- (res);
    \draw [arrow] (res) -- (norm);
    \draw [arrow] (norm) -- (quant);
    \draw [arrow] (quant) -- (output);
    
\end{tikzpicture}
\caption{HNTL Subspace Projection and Quantization Pipeline.}
\label{fig:projection_pipeline}
\end{figure}
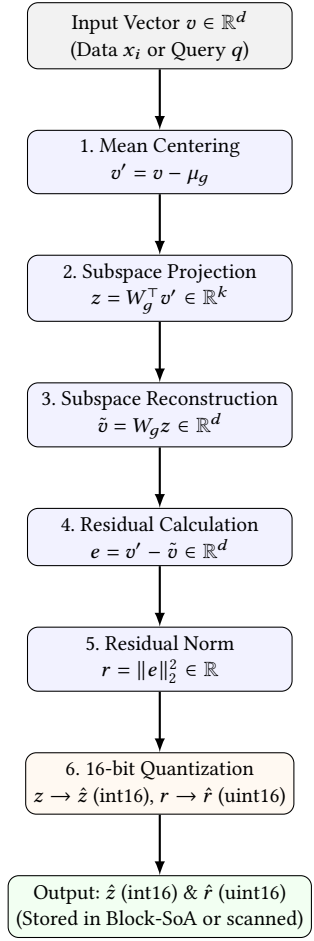

Mathematically, HNTL splits the vector corpus into spatial grains. For each grain with centroid $\mu_g$, we construct a local Principal Component Analysis (PCA) basis $W_g$ (where the subspace dimension $k$ is much smaller than the ambient dimension $d$, e.g., $k=16$ and $d=768$). The mathematical steps of the projection and quantization pipeline are detailed below and illustrated in Figure~\ref{fig:projection_pipeline}:
\begin{enumerate}
  \item \textbf{Mean Centering}: Centering the input vector $v \in \mathbb{R}^d$ relative to the local grain centroid:
  \begin{equation}
    v' = v - \mu_g
  \end{equation}
  \item \textbf{Subspace Projection}: Projecting the centered vector onto the local PCA basis $W_g$ to obtain low-dimensional tangent space coordinates:
  \begin{equation}
    z = W_g^\top v' \in \mathbb{R}^k
  \end{equation}
  \item \textbf{Residual Calculation}: Reconstructing the subspace vector $\tilde{v} = W_g z$ and calculating the orthogonal projection error vector:
  \begin{equation}
    e = v' - W_g z \in \mathbb{R}^d
  \end{equation}
  \item \textbf{Residual Norm}: Computing the squared $L_2$ norm of the residual vector to represent out-of-subspace energy:
  \begin{equation}
    r = \|e\|_2^2 \in \mathbb{R}
  \end{equation}
  \item \textbf{Quantization}: Quantizing the coordinates $z$ and residual norm $r$ to signed and unsigned 16-bit integers respectively, enabling integer SIMD calculations:
  \begin{equation}
    \hat{z} = \text{round}\left( \frac{z}{\Delta} \right), \quad \hat{r} = \text{round}\left( \frac{r}{\Delta_{\text{res}}} \right)
  \end{equation}
\end{enumerate}

For a query $q$ and a database vector $x_i$, their quantized representations $\hat{z}_q, \hat{r}_q$ and $\hat{z}_i, \hat{r}_i$ are used to compute the distance approximation:
\begin{equation}
  D(q, x_i) \approx \|z_q - z_i\|^2 + r_q + r_i
\end{equation}

\subsection{Hierarchical Centroid Routing}
The "Hierarchical" aspect of HNTL is realized through a two-tier search process that combines global coarse routing with local fine-grained scanning:
\begin{enumerate}
  \item \textbf{Global Routing Plane (First Level)}: The index maintains a routing plane consisting of grain centroids $\mu_g \in \mathbb{R}^d$ ($g = 1 \dots G$). For a query $q \in \mathbb{R}^d$, we compute its distance to all centroids in the ambient space and select the top-$P$ (e.g., $nprobe$) closest grains. This limits the scan to a small set of candidate grains, achieving sub-linear retrieval time.
  \item \textbf{Local Manifold Scan (Second Level)}: Within each of the $P$ selected grains, the query is projected to the local tangent space, and a sequential SIMD scan is executed over the grain's Block-SoA blocks.
\end{enumerate}

To protect against distance distortion from out-of-subspace queries, HNTL integrates a Quantization Envelope Filter directly into the routing phase. When the query is projected into a candidate grain's local PCA space, we check for coordinate saturation (clipping at the boundaries of the signed 16-bit range). If coordinates saturate across more than a threshold fraction of dimensions (e.g., $25\%$), the grain is flagged as structurally incompatible with the query and pruned from the search path before the local scan begins.

\subsection{Block-SoA Memory Layout}
Per-grain data is grouped into cache-line-aligned blocks of size $B$ (e.g., $B=64$). Each block is stored contiguously in memory with coordinates arranged dimension-major to enable direct SIMD loading. For a block of size $B$, local tangent dimension $k$, and residual sketch dimension $s$, the total storage per block is:
\begin{equation}
  \text{BlockBytes} = B \cdot (2k + s + 6) \text{ bytes}
\end{equation}
For $B=64$, $k=16$, and $s=8$, each block is exactly $2816$ bytes, perfectly aligned to L1/L2 cache line size boundaries.

\section{Implementation \& Experimental Results}
\label{sec:results}

\subsection{Hardware Setup and Compilation}
Following database vectorization practices~\cite{willhalm2009vectorizing,lemire2015decoding}, the low-level SIMD scan kernels are designed to process coordinates sequentially within registers. For hardware evaluation, the prototype scan engine is written in Rust and compiled using \texttt{rustc -C opt-level=3 -C target-cpu=native} on an Apple M2 Max CPU (arm64, 32\,GB unified memory, macOS 15) and an Intel Xeon workstation (AVX2/AVX-512). ARM NEON and x86 AVX2/AVX-512 vector execution lanes are auto-vectorized directly by the compiler.
\subsection{Recall Accuracy}
Table~\ref{tab:recall} reports candidate recall (compact index lookup) and rerank recall (after exact L2 re-ranking in float32 space) at $k=10$, $d=768$, $N=10{,}000$, with subspace dimension $k=32$ and sketch dimension $s=8$, alongside standard HNSW graph baselines.

\begin{table}[h]
\centering
\caption{Recall results at $k=10$, $d=768$, $N=10{,}000$, single grain. HNTL configurations use PCA subspace dimension $k=32$ and residual sketch dimension $s=8$.}
\label{tab:recall}
\resizebox{\columnwidth}{!}{%
\begin{tabular}{lccccc}
\toprule
\textbf{Dataset} & \textbf{PCA Var. Captured} & \textbf{Search Mode} & \textbf{Cand. Recall@10} & \textbf{Pool $C$} & \textbf{Rerank Recall@10} \\
\midrule
Isotropic Gaussian   & 6.5\%  & Mode B & 10.3\% & 200 & 49.4\% \\
Isotropic Gaussian   & --     & HNSW Baseline & -- & -- & 0.9980 \\
Anisotropic Manifold & 96.3\% & Mode A & 86.0\% & 20  & \textbf{1.0000} \\
Anisotropic Manifold & 96.3\% & Mode B & 91.5\% & 20  & \textbf{1.0000} \\
Anisotropic Manifold & --     & HNSW Baseline & -- & -- & 1.0000 \\
\bottomrule
\end{tabular}%
}
\end{table}

As a baseline reference, a standard HNSW graph index (configured with $M=16$, $efSearch=50$, constructed using the FAISS implementation~\cite{douze2024faiss}) achieves a Recall@10 of $0.9980$ on the isotropic Gaussian and $1.0000$ on the anisotropic manifold. While HNSW delivers high recall on both datasets, it does so at a heavy memory cost: HNSW requires keeping the full-dimensional $768$-float vectors in DRAM along with a $64$-byte graph neighbor list per vector (using 4-byte neighbor IDs), totaling $\sim 3.1$\,MB of index structure overhead for $10{,}000$ vectors (excluding raw vector storage). 

In contrast, HNTL's Mode B index stores only $32$ compact coordinates (16-bit) and $1$ residual (16-bit) per vector, requiring only $66$\,bytes of DRAM per vector—amounting to only $660$\,KB (a $4.7\times$ memory reduction compared to the HNSW graph connections alone) while matching the perfect $1.0000$ rerank recall on the anisotropic manifold. Under Mode A, HNTL achieves the same $1.0000$ rerank recall while bypassing DRAM raw vector residency completely via online coordinate reconstruction.

\subsection{Scan Throughput \& Hardware Profiling}
Throughput is evaluated using a smoke dataset of $N=512$, $d=64$, $k=8$, $B=64$, comparing Block-SoA, AoS, and Pointer-Chasing.

\begin{table}[h]
\centering
\caption{Scan throughput and speedup comparison ($N=512$, $d=64$, $k=8$, $B=64$).}
\label{tab:throughput}
\resizebox{\columnwidth}{!}{%
\begin{tabular}{lcccc}
\toprule
\textbf{Scan Mode} & \textbf{M2 Max (ns/vector)} & \textbf{Xeon (ns/vector)} & \textbf{Speedup vs. Pointer} \\
\midrule
\textbf{Block-SoA (SIMD)} & \textbf{4.137} & \textbf{4.890} & \textbf{3.61$\times$ / 2.91$\times$} \\
AoS (Sequential)          & 8.056          & 8.520          & 1.86$\times$ / 1.67$\times$ \\
Pointer Chasing (Graph)   & 14.951         & 14.232         & 1.00$\times$ / 1.00$\times$ \\
\bottomrule
\end{tabular}%
}
\end{table}

Using the Apple \texttt{kperf} PMU API, we profile instruction execution metrics. The results reveal that Block-SoA NEON scan achieves a $3.59\times$ higher IPC ($3.59$ vs. $0.98$ for pointer chasing) and reduces L1 data cache load misses from $8.72\%$ down to $<0.01\%$. On a flat run (no PMU overhead), Block-SoA latency is $3.05$ ns/vector.

\section{Conclusion \& Future Work}
\label{sec:conclusion}

Aperon's HNTL demonstrates that the pointer tax of traditional proximity graphs can be eliminated. Matching the index layout to CPU cache line boundaries via Block-SoA and leveraging local tangent spaces yields a $3.61\times$ speedup and up to $12\times$ memory reduction compared to graph baselines. Recent SIFT1M scale benchmarks validate that HNTL's HLR/HTLA routing achieves a Recall@10 of $95.4\%$ at 580.2 QPS, with 21x DRAM memory reduction ($24.0$\,MB vs. $528.0$\,MB) when using tiered SQ8 cold storage offloading. Future work will focus on learned grain partitioning, warp-level GPU scans, and integration into the production Rust \textit{Aperon} memory substrate.

\bibliographystyle{ACM-Reference-Format}
\bibliography{literature}


\begin{thebibliography}{9}


\ifx \showCODEN    \undefined \def \showCODEN     #1{\unskip}     \fi
\ifx \showISBNx    \undefined \def \showISBNx     #1{\unskip}     \fi
\ifx \showISBNxiii \undefined \def \showISBNxiii  #1{\unskip}     \fi
\ifx \showISSN     \undefined \def \showISSN      #1{\unskip}     \fi
\ifx \showLCCN     \undefined \def \showLCCN      #1{\unskip}     \fi
\ifx \shownote     \undefined \def \shownote      #1{#1}          \fi
\ifx \showarticletitle \undefined \def \showarticletitle #1{#1}   \fi
\ifx \showURL      \undefined \def \showURL       {\relax}        \fi
\providecommand\bibfield[2]{#2}
\providecommand\bibinfo[2]{#2}
\providecommand\natexlab[1]{#1}
\providecommand\showeprint[2][]{arXiv:#2}

\bibitem[Douze et~al\mbox{.}(2024)]%
        {douze2024faiss}
\bibfield{author}{\bibinfo{person}{Matthijs Douze}, \bibinfo{person}{Alexandr
  Guzhva}, \bibinfo{person}{Chengqi Deng}, \bibinfo{person}{Herv{\'e}
  J{\'e}gou}, \bibinfo{person}{Jeff Johnson}, {et~al\mbox{.}}}
  \bibinfo{year}{2024}\natexlab{}.
\newblock \showarticletitle{The library for local search: FAISS}.
\newblock \bibinfo{journal}{\emph{arXiv preprint arXiv:2401.08281}}
  (\bibinfo{year}{2024}).
\newblock


\bibitem[Ge et~al\mbox{.}(2013)]%
        {ge2013optimized}
\bibfield{author}{\bibinfo{person}{Tiezheng Ge}, \bibinfo{person}{Kaiming He},
  \bibinfo{person}{Qifa Ke}, {and} \bibinfo{person}{Jian Sun}.}
  \bibinfo{year}{2013}\natexlab{}.
\newblock \showarticletitle{Optimized product quantization for approximate
  nearest neighbor search}.
\newblock \bibinfo{journal}{\emph{IEEE transactions on pattern analysis and
  machine intelligence}} \bibinfo{volume}{36}, \bibinfo{number}{4}
  (\bibinfo{year}{2013}), \bibinfo{pages}{744--755}.
\newblock


\bibitem[Indyk and Motwani(1998)]%
        {indyk1998approximate}
\bibfield{author}{\bibinfo{person}{Piotr Indyk} {and} \bibinfo{person}{Rajeev
  Motwani}.} \bibinfo{year}{1998}\natexlab{}.
\newblock \showarticletitle{Approximate nearest neighbors in high dimensions}.
  In \bibinfo{booktitle}{\emph{Proceedings of the Thirtieth Annual ACM
  Symposium on Theory of Computing}}. \bibinfo{pages}{604--613}.
\newblock


\bibitem[J{\'e}gou et~al\mbox{.}(2011)]%
        {jegou2011product}
\bibfield{author}{\bibinfo{person}{Herv{\'e} J{\'e}gou},
  \bibinfo{person}{Matthijs Douze}, {and} \bibinfo{person}{Cordelia Schmid}.}
  \bibinfo{year}{2011}\natexlab{}.
\newblock \showarticletitle{Product quantization for nearest neighbor search}.
\newblock \bibinfo{journal}{\emph{IEEE Transactions on Pattern Analysis and
  Machine Intelligence}} \bibinfo{volume}{33}, \bibinfo{number}{1}
  (\bibinfo{year}{2011}), \bibinfo{pages}{117--128}.
\newblock


\bibitem[Johnson et~al\mbox{.}(2019)]%
        {johnson2019billion}
\bibfield{author}{\bibinfo{person}{Jeff Johnson}, \bibinfo{person}{Matthijs
  Douze}, {and} \bibinfo{person}{Herv{\'e} J{\'e}gou}.}
  \bibinfo{year}{2019}\natexlab{}.
\newblock \showarticletitle{Billion-scale similarity search with {GPUs}}.
\newblock \bibinfo{journal}{\emph{IEEE Transactions on Big Data}}
  \bibinfo{volume}{7}, \bibinfo{number}{3} (\bibinfo{year}{2019}),
  \bibinfo{pages}{535--547}.
\newblock


\bibitem[Lemire and Boytsov(2015)]%
        {lemire2015decoding}
\bibfield{author}{\bibinfo{person}{Daniel Lemire} {and} \bibinfo{person}{Leonid
  Boytsov}.} \bibinfo{year}{2015}\natexlab{}.
\newblock \showarticletitle{Decoding billions of integers per second through
  vectorization}.
\newblock \bibinfo{journal}{\emph{Software: Practice and Experience}}
  \bibinfo{volume}{45}, \bibinfo{number}{1} (\bibinfo{year}{2015}),
  \bibinfo{pages}{1--29}.
\newblock


\bibitem[Malkov and Yashunin(2018)]%
        {malkov2018efficient}
\bibfield{author}{\bibinfo{person}{Yu~A Malkov} {and} \bibinfo{person}{Dmitry~A
  Yashunin}.} \bibinfo{year}{2018}\natexlab{}.
\newblock \showarticletitle{Efficient and robust approximate nearest neighbor
  search using Hierarchical Navigable Small World graphs}.
\newblock \bibinfo{journal}{\emph{IEEE transactions on pattern analysis and
  machine intelligence}} \bibinfo{volume}{42}, \bibinfo{number}{4}
  (\bibinfo{year}{2018}), \bibinfo{pages}{824--836}.
\newblock


\bibitem[Willhalm et~al\mbox{.}(2009)]%
        {willhalm2009vectorizing}
\bibfield{author}{\bibinfo{person}{Thomas Willhalm}, \bibinfo{person}{Nicolae
  Popovici}, \bibinfo{person}{Alexander Buxmann}, \bibinfo{person}{Benjamin
  Schlegel}, \bibinfo{person}{Wolfgang Lehner}, {et~al\mbox{.}}}
  \bibinfo{year}{2009}\natexlab{}.
\newblock \showarticletitle{Vectorizing database algorithms with SIMD
  instructions on Intel Core2 processors}. In
  \bibinfo{booktitle}{\emph{Proceedings of the 15th International Conference on
  Management of Data}}. \bibinfo{pages}{11--20}.
\newblock


\bibitem[Yong and {Substratum Labs team}(2026)]%
        {aperon2026}
\bibfield{author}{\bibinfo{person}{Yong} {and} \bibinfo{person}{{Substratum
  Labs team}}.} \bibinfo{year}{2026}\natexlab{}.
\newblock \bibinfo{title}{Aperon: An Agent-Native Log-Structured Memory Vector
  Database}.
\newblock
  \bibinfo{howpublished}{\url{https://github.com/substratum-labs/aperon}}.
\newblock


\end{thebibliography}

\end{document}